\documentclass [a4paper]{article}

\usepackage{amsmath,amssymb, theorem}
\newtheorem{prop}{Proposition}
\newtheorem{theorem}{Theorem}
\newtheorem{lemma}{Lemma}
\newtheorem{coro}{Corollary}
\theorembodyfont{\rm}

\newtheorem{rem}{Remark}

\theorembodyfont{\rm}
\newtheorem{remark}{Remark}
\def\Fe{\mathcal F}
\def\Ee{\mathcal E}
\def\Me{\mathcal M}

\def\Ae{\mathcal A}
\def\Se{\mathcal S}

\def\Be{\mathcal B}
\def\Ha{\mathcal H}
\def\Ka{\mathcal K}

\def\Pe{\mathcal P}
\def\<{\langle}
\def\>{\rangle}
\def\qed{{\hfill $\square$}}
\def\Tr {{\rm Tr}\,}
\def\supp {{\rm supp}\,}

\title{Comparison of  quantum binary experiments}
\author{Anna Jen\v cov\'a\thanks{Supported by   the grants 
VEGA 2/0032/09 and meta-QUTE ITMS 26240120022.}
\\
Mathematical Institute, Slovak Academy of Sciences,\\
\v {S}tef\'{a}nikova 49, 814 73 Bratislava, Slovakia\\ e-mail: 
jenca@mat.savba.sk}
\date{}

\begin{document}
\maketitle
\begin{abstract} A quantum binary experiment consists of 
 a pair of density operators on a finite dimensional Hilbert space. 
An experiment $\Ee$ is called $\epsilon$-deficient with respect to another 
experiment $\Fe$ if, up to $\epsilon$, 
 its risk functions are not worse than the risk functions of $\Fe$, with respect to all statistical decision problems. It is
 known in the theory of classical statistical experiments that 1.   
 for  pairs of 
probability distributions, one can restrict to testing problems in the definition of deficiency
and 2. that 0-deficiency is  a necessary and sufficient condition 
for
 existence of a stochastic mapping that maps one pair onto the other.
We show that in the quantum case, the property 1. holds precisely if $\Ee$ 
consist of commuting densities. As for property 2., we show that if $\Ee$ is 0-deficient with respect to $\Fe$, then there exists a completely positive mapping that maps $\Ee$ onto $\Fe$, but it is not necessarily trace preserving.

\end{abstract}

\section{Introduction}

In classical statistics, a statistical experiment is a parametrized family of probability distributions 
on a sample space $(X,\Sigma)$.  
The theory of experiments and their comparison was introduced by Blackwell \cite{blackwell} and further 
developed by many authors, e.g. Torgersen, \cite{torgersen,torgersenb}.
Most of the results needed here can be found in \cite{strasser}.

For our purposes, a classical \emph{statistical experiment} $\Ee=(X,\{p_\theta,\ \theta\in \Theta\})$ is a parametrized set  of
probability distributions $p_\theta, \theta\in\Theta$ over a finite set $X$, where $\Theta$ is a finite set of parameters.
This can be interpreted as follows: $X$ is a set of possible outcomes $x\in X$ of some experiment, each occurring with
probability $p(x)$, where $p$ is a member of the parametrized family $\{p_\theta\}$, but the  value of the parameter is not
known.
After observing $x$, a decision $d$ is chosen from a finite set $D$ of possible decisions, with some probability $\mu(x,d)$.
The function  $\mu: X\times D\to [0,1]$ is called the  \emph{decision function}.
It   is clear that a decision function is a Markov kernel (or a stochastic matrix), that is, $d\mapsto \mu(x,d)$ is a probability
 distribution for all $x\in X$. 

A \emph{loss function} $W: \Theta\times D\to \mathbb R^+$ represents the loss suffered if $d\in D$ is chosen
 and the true value of the parameter is $\theta$. The \emph{risk}, or the average loss  of the decision procedure $\mu$ 
 when the true value is $\theta$ is computed as
 \[
R_\Ee(\theta,W,\mu)=\sum_{x,d} W_\theta(d)\mu(x,d)p_\theta(x)
 \]
The couple $(D,W)$ is called a \emph{decision problem}. If $D$ consists of two points, then the decision problems $(D,W)$ are 
precisely the problems of hypothesis testing.

Let $\Fe$ be another experiment with the same set of parameters, then its ''informative value'' can be compared to that of $\Ee$ 
by comparing their risk functions for all decision problems. This leads to the definitions of $(k,\epsilon)$-deficiency and $\epsilon$-deficiency,  see Section \ref{sec:def}.
One of the most important results of the theory is the following \emph{randomization criterion}:

\begin{theorem} Let $\Ee=(X,\{p_\theta, \theta\in \Theta\})$ and $\Fe=(Y,\{q_\theta,\theta\in \Theta\})$ be two experiments. Then $\Ee$ is $\epsilon$-deficient
 with respect to $\Fe$ if and only if there is a Markov kernel $\lambda: X\times Y\to [0,1]$ such that 
 \[
 \|\lambda(p_\theta)-q_\theta\|_1\le 2\epsilon
 \]
 where $\lambda(p)= \sum_x \lambda(x,y)p(x)$.
 \end{theorem} 

For $\epsilon =0$, this is the Blackwell-Sherman-Stein Theorem, \cite{blackwell, sherman,stein}. For general $\epsilon$ it was proved in \cite{torgersen}.

If $\Theta$ consists of two points, then the  experiment is called \emph{binary}. In this case, $\epsilon$-deficiency is equivalent to $(2,\epsilon)$-deficiency \cite{torgersen}, which means that such experiments
can  be compared by considering only the risk functions of hypothesis testing problems.

The development of the quantum version of comparison of statistical experiments was started recently by several authors, \cite{shmaya,buscemi,matsumoto}.  A quantum statistical experiment is a set of density operators on a Hilbert space, mostly of finite 
dimension.  
Some versions of the randomization criterion, resp. the Blackwell-Sherman-Stein Theorem were obtained, in particular, conditions 
were found 
for existence of a trace preserving completely positive map that maps one experiment onto the other. It was conjectured in 
\cite{shmaya} that  the existence of  such positive (but not necessarily completely positive) trace preserving map is equivalent
 to 0-deficiency. A weaker form of this was obtained in \cite{buscemi}, where the notion of a
 \emph{statistical morphism} was introduced. The (even weaker) notion of a \emph{$k$-statistical morphism} was considered in \cite{matsumoto}.

The present paper reviews some of the results of \cite{buscemi} and \cite{matsumoto}, with focus on 
the problem of comparison of binary experiments. As an extension of \cite{matsumoto}, 
we
prove that  $(2,\epsilon)$-deficiency and $\epsilon$- deficiency of a quantum experiment 
$\Ee$ with respect to another quantum experiment $\Fe$ are equivalent for any $\Fe$
 precisely if the experiment $\Ee$ is abelian, that is, all density matrices $\rho_\theta$ commute. 
 Moreover, we use the results in \cite{wolf} to show that any
 $k$- statistical morphism can be extended to a map that is completely positive, but not trace preserving in general.

\section{Quantum statistical experiments}\label{sec:qse}

 Let $\Ha$ be a finite dimensional Hilbert space and let $\Ae\subseteq B(\Ha)$ be a $C^*$-algebra. Let 
 $\Se(\Ae)$ denote the set of density operators in  $\Ae$.
A (quantum) statistical experiment $\Ee$ consists of $\Ae$ and a  family
$\{\rho_\theta,\theta\in \Theta\}\subset \Se(\Ae)$,  which is written as $\Ee=(\Ae,\{ \rho_\theta,\theta\in \Theta\})$.
 Throughout the paper, we suppose that  $\Theta$ is a finite set.

The family $\{\rho_\theta,\theta\in \Theta\}$ represents our knowledge of the state of the quantum system represented by $\Ae$:
it is known that  this  family contains the state of the system  but the true value of $\theta$ is not known.

Let   $(D,W)$ be a decision problem.
The decision is made by a measurement on $\Ae$ with values in $D$. Any such measurement is given by a positive operator valued measure (POVM) $M: D\to \Ae$,
 that is, a  collection of operators $M=\{M_d, d\in D\}\subset \Ae^+$ such that $\sum_dM_d=I$. If all $M_d$ are projections,
  we say that $M$ is a projection valued measure (PVM).
We will denote the set of all  measurements by $\Me(D,\Ee)$.

 Note that any POVM defines  a
 positive trace preserving map $M:\Ae\to \Fe(D)$, where $\Fe(D)$ is  the $C^*$-algebra of all functions $D\to \mathbb C$.
 The map is given by
 \[
 M(a)(d)=\Tr M_da, \qquad a\in \Ae, \ d\in D
 \]
 and any positive trace preserving map $\Ae\to \Fe(D)$ is obtained in this way. Moreover, we define the  map $\hat M:\Fe(D)\to \Ae$ by
\[
\hat M(f)=\sum_d f(d)(\Tr M_d)^{-1}M_d,\qquad f\in \Fe(D).
\]
Then $\hat M$ is again positive and trace preserving.
 Since $\Fe(D)$ is abelian, both $M$ and $\hat M$ are also completely positive,
 \cite{paulsen}.

As it was pointed out in \cite{buscemi}, the set of quantum experiments contains 
the set of classical experiments and
these correspond precisely to abelian experiments, that is, experiments such that all densities in the family
$\{\rho_\theta,\theta\in\Theta\}$ commute. Indeed, let $\Ee$ be abelian and let
$\mathcal C$ be the subalgebra generated by $\{\rho_\theta,\theta\in\Theta\}$. Then $\mathcal C$ is generated by a PVM $P$ 
concentrated on a
finite set $X$ and we have the classical experiment $(X,\{p_\theta:=P(\rho_\theta),\theta\in \Theta\})$. Conversely,
let $(Y,\{q_\theta,\theta\in \Theta\})$ be any classical experiment with $|Y|\le \dim(\Ha)$ and let $Q:Y\to \Ae$ be any PVM,
 then $(\Ae,\{ \hat Q(q_\theta), \theta\in \Theta\})$ defines an abelian quantum experiment.
  It is clear that $p_\theta=P(\rho_\theta)$ and $\rho_\theta=\hat P(p_\theta)$, $\theta\in \Theta$, so that $\Ee$ and 
  $(X,\{p_\theta\})$ are mapped onto each other by completely positive trace preserving maps. In particular, the experiments 
  are equivalent 
   in the sense defined below.

\section{Deficiency}\label{sec:def}

Let $\Ee$ be an experiment and let $(D,W)$ be a decision problem.
The \emph{risk} of the decision procedure $M\in \Me(D,\Ee)$ at $\theta$ is computed as \cite{holevo}
\[
R_{\mathcal E} (\theta, W,M)= \sum_{d\in D}M(\rho_\theta)(d)W_\theta(d)=\sum_d W_\theta(d)\Tr \rho_\theta M_d
\]

Let now $\Fe=(\Be,\{\sigma_\theta, \theta\in \Theta\})$ be another experiment, with
$\Be\subset B(\Ka)$ for a finite dimensional Hilbert space $\Ka$ and  with the same parameter set.
Let $k\in \mathbb N$, $D_k:=\{0,\dots,k-1\}$ and let $\epsilon\ge 0$. We say that $\Ee$ is \emph{$(k,\epsilon)$-deficient with respect to
$\Fe$}, in notation $\Ee \ge_{k,\epsilon}\Fe$, if for every decision problem $(D_k,W)$ (equivalently, for all decision problems
$(D,W)$ with $|D|=k$) and every $N\in \Me(D_k,\Fe)$, there is
some $M\in\Me(D_k,\Ee)$ such that
\[
R_\Ee(\theta,W,M)\le R_\Fe(\theta,W, N)+\epsilon\|W_\theta\|,\qquad \theta\in \Theta
\]
where  $\|W_\theta\| =\sup_{x\in D_k} W_\theta(x)$.
We say that $\Ee$ is \emph{$\epsilon$-deficient with respect to $\Fe$}, $\Ee\ge_{\epsilon}\Fe$,  if it is $(k,\epsilon)$-deficient for all $k\in \mathbb N$.

The relation $\le_0$ defines a preorder on the set of all experiments. If we have $\Ee\ge_0 \Fe$ and simultaneously
$\Fe\ge_0 \Ee$, then we say that $\Ee$ and $\Fe$ are \emph{equivalent}, $\Ee\sim\Fe$. The equivalence relation
$\Ee\sim_{k}\Fe$ is defined similarly, and $\Ee$ and $\Fe$ are called \emph{$k$-equivalent}.

The Theorem \ref{thm:equiv} below   (apart from (iii)) was proved in \cite[Theorem 5]{matsumoto} in a more general setting.  
We give the proof in our simpler case, just for the convenience of the reader.

The most important ingredient of the proof is the \emph{minimax theorem}, which can be found in 
\cite{strasser}.

\begin{theorem}\label{thm:equiv} 
Let $\Ee=(\Ae,\{\rho_\theta,\theta\in \Theta\})$ and $\Fe=(\Be,\{\sigma_\theta,\theta\in\Theta\})$
be two
experiments with the same parameter set $\Theta$, $|\Theta|<\infty$. Let $k\in \mathbb N$, $\epsilon \ge 0$.
The following are equivalent.
\begin{enumerate}
\item[(i)] $\Ee\ge_{k,\epsilon}\Fe$
\item[(ii)] For every loss function $W:\Theta\times D_k\to \mathbb R^+$,
\[
\min_{M\in \Me(D_k,\Ee)}\sum_\theta R_\Ee(\theta,W,M)\le \min_{N\in \Me(D_k,\Fe)}\sum_\theta R_\Fe(\theta, W,N) + \epsilon \|W\|
\]
where $\|W\|=\sum_\theta \|W_\theta\|$.
\item[(iii)] For every loss function $W:\Theta\times D_k\to \mathbb R^+$,
\[
\max_{M\in \Me(D_k,\Ee)}\sum_\theta R_\Ee(\theta,W,M)\ge \max_{N\in \Me(D_k,\Fe)}\sum_\theta R_\Fe(\theta, W,N) - \epsilon \|W\|
\]
\item[(iv)] For every $N\in \Me(D_k,\Fe)$ there is some $M\in \Me(D_k,\Ee)$ such that
\[
\|M(\rho_\theta)-N(\sigma_\theta)\|_1\le 2\epsilon, \qquad \forall \theta\in \Theta
\]

\end{enumerate}
\end{theorem}

{\it Proof.}
Suppose (i), then for any $N\in \Me(D_k,\Fe)$, there is some $M\in \Me(D_k,\Ee)$ such that
\[
\sum_\theta R_\Ee(\theta,W,M)\le \sum_\theta R_\Fe(\theta,W,N)+\epsilon\|W\|,
 \]
this implies (ii).

Suppose (ii) and let $W:\Theta\times D_k\to \mathbb R^+$ be a loss function. Then $\tilde W:\Theta\times D_k\to \mathbb R^+$
 given by $\tilde W_\theta=\|W_\theta\|-W_\theta$ is a loss function with $\|\tilde W\|\le\|W\|$.
Since $R_\Ee(\theta,\tilde W,M)=\|W_\theta\|-R_\Ee(\theta,W,M)$ and similarly for $R_\Fe$, we have (ii) implies (iii).

Suppose (iii), and let $N\in \Me(D_k,\Fe)$. Then for every loss function $W$, we have
\[
\max_{M \in \Me(D_k,\Ee)}\sum_\theta R_\Ee(\theta, W,M)\ge \sum_\theta R_\Fe(\theta,W,N)-\epsilon\|W\|,
\]
and this implies that
\[
\sup_{W,\|W\|\le 1}\min_{M \in \Me(D_k,\Ee)}\sum_\theta (R_\Fe(\theta, W,N)- R_\Ee(\theta,W,M))\le \epsilon
\]
The set $\Me=\Me(D_k,\Ee)$ is compact and obviously convex and the set $\mathcal W$ of all loss functions $W$ with $\|W\|\le 1$ 
is convex as well. Moreover, the function $(M, W)\mapsto \sum_\theta (R_\Fe(\theta, W,N)- R_\Ee(\theta,W,M))$ is linear
 in both arguments, hence the minimax theorem applies and we get
\begin{eqnarray*}
\epsilon \ge \min_{M\in \Me}\sup_{W\in \mathcal W} \sum_\theta (R_\Fe(\theta, W,N)- R_\Ee(\theta,W,M))\\
=\min_{M\in \Me}\sup_{W\in \mathcal W} \sum_{\theta,d} W_\theta(d)(N(\sigma_\theta)(d)-M(\rho_\theta)(d))
\end{eqnarray*}
Let $\Pe(\Theta)$ be the set of all probability measures on $\Theta$ and let $p\in \Pe(\Theta)$.
For $M\in \Me$ fixed,  let $W$ be given by
\[
W_\theta(x)=\left\{ \begin{array}{cc} p(\theta) & \mbox{if }
N(\sigma_\theta)(x)-M(\rho_\theta)(x)>0\\
0 & \mbox{otherwise} \end{array}\right.
\]
Then $W\in \mathcal W$, so that we get
\begin{eqnarray*}
\epsilon \ge  \min_{M\in \Me}\sum_\theta\sum_{x\in D_k} W_\theta(x)(N(\sigma_\theta)(x)-M(\rho_\theta)(x))\\
=\min_{M\in \Me}\sum_\theta p(\theta)\frac 12 \|N(\sigma_\theta)-M(\rho_\theta)\|_1
\end{eqnarray*}
Since this holds for any $p\in \Pe(\Theta)$, we have obtained
\[
\sup_{p\in \Pe(\Theta)}\min_{M\in \Me}\sum_\theta p(\theta)\|M(\rho_\theta)-N(\sigma_\theta)\|_1\le 2\epsilon
\]
The set  $\Pe(\Theta)$ is convex and  the function
$\Me\times \Pe(\Theta)\to \mathbb R$, given by $(M,p)\mapsto \sum_\theta p(\theta)\|M(\rho_\theta)-N(\sigma_\theta\|_1$
 is convex in $M$ and concave (linear) in $p$. Hence the minimax theorem applies again and we have
 \[
\min_M\sup_p\|M(\rho_\theta)-N(\sigma_\theta)\|_1=\sup_p\min_M\sum_\theta p(\theta)\|M(\rho_\theta)-N(\sigma_\theta)\|_1\le 2\epsilon
 \]
which clearly implies (iv), by taking the probability measures concentrated in $\theta\in \Theta$.

Suppose (iv) and let $N\in \Me(D_k,\Fe)$. Let $M\in \Me(D_k,\Ee)$ be chosen for $N$ by (iv). Then for any loss function $W$,
\begin{eqnarray*}
R_\Ee(\theta,W,M)-R_\Fe(\theta,W,N)=\sum_{x\in D_k} W_\theta(x) (M(\rho_\theta)(x)-N(\sigma_\theta)(x))\\
\le
\frac {\|W_\theta\|}2\|M(\rho_\theta)-N(\sigma_\theta)\|_1\le \epsilon\|W_\theta\|
\end{eqnarray*}
so that $\Ee\ge_{k,\epsilon}\Fe$.

\qed

The following Corollary is a generalization of the classical randomization criterion to the case when the experiment $\Fe$ is abelian.  In the case that $\epsilon=0$,
it was proved in \cite{buscemi}.

\begin{coro}\label{coro:abelian} Let $\Ee=(\Ae,\{\rho_\theta,\theta\in \Theta\})$ and let
$\Fe=(\Be,\{\sigma_\theta, \theta\in \Theta\})$ be
 abelian.  Then  $\Ee \ge_\epsilon \Fe$
if and only if there is a
completely positive trace preserving map $T:\Ae\to \Be$ such that
\[
\|T(\rho_\theta)-\sigma_\theta\|_1\le 2\epsilon,\quad \theta\in \Theta
\]
\end{coro}

{\it Proof.} Let $(X,\{p_\theta,\theta\in \Theta\})$ be a classical experiment equivalent to $\Fe$ and let
$P=(P_1,\dots,P_m)$ be the PVM such that  $P(\sigma_\theta)=p_\theta$, $\theta\in \Theta$.
Suppose $\Ee\ge_\epsilon \Fe$, then   $P\in \Me(X,\Fe)$ and  by Theorem \ref{thm:equiv} (iv), there is some
$M\in \Me(X,\Ee)$ such that
\[
\|M(\rho_\theta)-P(\sigma_\theta)\|_1=\|M(\rho_\theta)-p_\theta\|_1\le 2\epsilon
\]
Put $T=\hat P\circ M$, then $T:\Ae\to \Be_0\subseteq \Be$ is positive and trace preserving, where $\Be_0$ is the abelian subalgebra
 generated by $P$. Hence  $T$ is also completely positive. Moreover,
\[
\|T(\rho_\theta)-\sigma_\theta\|_1=\|\hat P(M(\rho_\theta)-p_\theta)\|_1\le
\|M(\rho_\theta)-p_\theta\|_1\le 2\epsilon
\]

For the converse, let  $N\in \Me(D,\Fe)$ for any finite set $D$. Put
$Q=N\circ T$,  then $Q\in \Me(D,\Ee)$ and
\[
\|Q(\rho_\theta)-N(\sigma_\theta)\|_1=\|N(T(\rho_\theta)-\sigma_\theta)\|_1\le 2\epsilon
\]
By Theorem \ref{thm:equiv} (iv), this implies $\Ee\ge_\epsilon \Fe$.

\qed

\subsection{Deficiency w.r. to testing problems}

Let $(D_2, W)$ be a  decision problem. Then any $M\in \Me(D_2,\Ee)$ has the form $(M_0,I-M_0)$ for some
$0\le M_0\le I$ and the risk of $M$ is
\[
R_\Ee(\theta,M,W)=W_\theta(1)+ (W_\theta(0)-W_\theta(1))\Tr \rho_\theta M_0
\]

By Theorem \ref{thm:equiv} (iii), $\Ee\ge_{2,\epsilon}\Fe$ if and only if
\begin{equation}\label{eq:testsw}
\max_{\substack{ M_0\in \Ae,\\ 0\le M_0 \le 1}}\Tr \sum_\theta A_\theta \rho_\theta M_0\ge
\max_{\substack{N_0\in \Be,\\ 0\le N_0 \le 1}}\Tr \sum_\theta A_\theta \sigma_\theta N_0-\epsilon\|W\|
\end{equation}
for all loss functions $W$, where we denote $A_\theta:=W_\theta(0)-W_\theta(1)$. It is easy to see that
\begin{equation}\label{eq:testsa}
\max_{0\le M_0 \le 1}\Tr \sum_\theta A_\theta\rho_\theta M_0=
\Tr\left[\sum_\theta A_\theta \rho_\theta\right]^+=\frac 12 (\sum_\theta A_\theta
+\|\sum_\theta A_\theta \rho_\theta\|_1),
\end{equation}
here we used the equality $\Tr a^+=\frac12(\Tr a+\Tr |a|)$ 
 for a self adjoint element $a\in \Ae$.
\begin{theorem}\label{thm:testing}
$\Ee\ge_{2,\epsilon}\Fe$ if and only if
\[
\|\sum_\theta A_\theta \rho_\theta\|_1\ge \|\sum_\theta A_\theta \sigma_\theta\|_1
-2\varepsilon \sum_\theta |A_\theta|
\]
for any coefficients $A_\theta\in \mathbb R$.
\end{theorem}

{\it Proof.} Follows from (\ref{eq:testsw}) and (\ref{eq:testsa}). For the 'if' part,
put $A_\theta=
W_\theta(0)-W_\theta(1)$, we then have $\sum_\theta|A_\theta|\le \|W\|$. For the converse, let
$F_+:=\{ \theta, A_\theta>0\}$, $F_-:=\{\theta, A_\theta\le 0\}$ and put
$W_\theta(0)=\left\{ \begin{array}{cc} A_\theta & \mbox{if } \theta\in F_+\\
                                        0 &\mbox{otherwise}
					\end{array}\right.$,
 $W_\theta(1)=\left\{ \begin{array}{cc} -A_\theta & \mbox{if } \theta\in F_-\\
                                        0 &\mbox{otherwise}
					\end{array}\right.$.
Then $W$ is a loss function with $\|W\|=\sum_\theta |A_\theta|$.

\qed

\subsection{Deficiency and sufficiency}

Let $T:\Ae \to \Be$ be a completely positive trace preserving map. The experiment $\Fe=(\Be,\{T(\rho_\theta),\theta\in \Theta\})$
 is called a \emph{randomization} of $\Ee$.  If $N\in \Me(D,\Fe)$, then $T^*(N)\in \Me(D,\Ee)$ and it is clear that
 $T^*(N)$ has the same risks as $N$, hence $\Ee$ is 0-deficient  with respect to $\Fe$.

Suppose that in this setting, $\Fe$ is $k,0$-deficient with respect to $\Ee$, then we say that $T$ is \emph{$k$-sufficient } for
$\Ee$. If also $\Ee$ is  a randomization of $\Fe$, then we say that $T$ is \emph{sufficient} for $\Ee$, this definition of sufficiency was introduced in \cite{petz}. If $T$ is a restriction to a subalgebra $\Ae_0\subset \Ae$, then we say that 
$\Ae_0$ is $k$-sufficient resp. sufficient for $\Ee$, if $T$ is. If the experiments are abelian, then 
it follows by the randomization
criterion that  $T$ is sufficient if and only if it is $k$-sufficient for every $k\in \mathbb N$.
Moreover, for abelian binary experiments,  $T$ is sufficient if and only if it is 2-sufficient.
(In fact, the last statement hold for all classical statistical experiments \cite{strasser}.)

It is not clear if any of the above two statements holds for quantum  experiments.
The latter condition for binary experiments was investigated in \cite{ja_hypo}, for a subalgebra $\Ae_0$. It was shown that 
$\Ae_0$ is 2-sufficient if and only if it contains all projections $P_{t,+}$, $t\ge 0$ (see  Lemma \ref{lemma:qnp}) and that
this is equivalent to sufficiency in some cases. In particular:

\begin{theorem}\label{thm:ja_hypo} Let $\Ee=(\Ae,\{\rho_1,\rho_2\})$ be an experiment and let $\Ae_0\subseteq \Ae$ be an abelian
subalgebra. Then the following are equivalent.
\begin{enumerate}
\item [(i)] $\Ae_0$ is 2-sufficient.
\item [(ii)] $\Ae_0$ is sufficient.
\item[(iii)] $\Ae_0$ is sufficient and $\Ee$ is abelian.

\end{enumerate}

\end{theorem}

{\it Proof.} The equivalence of (i) and (ii) was proved in \cite[Thm. 5(2)]{ja_hypo}, (ii) $\implies$ (iii) follows from
\cite[Theorem 9.10]{ohypetz}. (iii) $ \implies$ (i) is obvious.

\qed

\section{Binary experiments}

Let $\Ee=(\Ae,\{\rho_1,\rho_2\})$ be a binary experiment. Note that we may 
suppose that $\rho_1+\rho_2$ is invertible, since $\Ee$ can be replaced by the experiment $(P\Ae P,\{\rho_1,\rho_2\})$,
 where $P=\supp (\rho_1+\rho_2)$ is the support projection of $\rho_1+\rho_2$.

Let us denote
\[
f_\Ee(t):= \max_{\substack{M\in \Ae,\\0\le M\le I}} \Tr (\rho_1-t\rho_2)M,\qquad t\in \mathbb R
\]
Then by (\ref{eq:testsa}),
\begin{equation}\label{eq:fe}
f_\Ee(t)=\Tr(\rho_1-t\rho_2)_+=\frac 12(\|\rho_1-t\rho_2\|_1+1-t)
\end{equation}
It is easy to see that Theorem \ref{thm:testing} for binary experiments has the following form.

\begin{theorem}\label{thm:binary} Let $\Ee=\{\Ae,\{\rho_1,\rho_2\})$ and $\Fe=(\Be,\{\sigma_1,\sigma_2\})$. Then the following are equivalent.
\begin{enumerate}
\item [(i)] $\Ee\ge_{2,\epsilon}\Fe$
\item[(ii)] $\|\rho_1-t\rho_2\|_1\ge \|\sigma_1-t\sigma_2\|_1 -2(1+t)\varepsilon$
for all $t\ge 0$.
\item [(iii)]  $f_\Ee(t)\ge f_\Fe(t)-(1+t)\epsilon$ for all $t\ge 0$.
\end{enumerate}
\end{theorem}

We will need some properties of the function $f_\Ee$. 
First, we state the  quantum version of the Neyman-Pearson lemma, \cite{helstrom, holevo}. For this, let us denote $P_{t,+}:={\rm supp}\, (\rho_1-t\rho_2)_+$ and $P_{t,0}={\rm ker}\,(\rho_1-t\rho_2)$ for $t\ge0$.

\begin{lemma}\label{lemma:qnp} We have $f_\Ee(t)=\Tr (\rho_1-t\rho_2)M$ for some $M\in \Ae$, $0\le M\le I$ if and only if
\[
M=P_{t,+}+X,\qquad 0\le X\le P_{t,0}
\]

\end{lemma}

The proof of the following lemma can be found in the Appendix.

\begin{lemma}\label{lemma:fe}
\begin{enumerate}
\item[(i)] $f_\Ee$ is continuous, convex and $f_\Ee(t)\ge \max\{1-t,0\}$, $t\in \mathbb R$.
\item[(ii)] $f_\Ee$ is nonincreasing in $\mathbb R$. Moreover,
$f_\Ee$ is analytic in $\mathbb R$ except some points $0\le t_1<\dots<t_l$, $l\le \dim(\Ha)$, where $f_\Ee$ is 
not differentiable.
These are exactly the points where $P_{t,0}\ne 0$.
\end{enumerate}
\end{lemma}

We will  denote $\mathcal T_\Ee:=\{t_1,\dots,t_l\}$ the set of points defined in 
(ii).

\subsection{Deficiency and 2-deficiency for binary experiments}

For classical binary experiments, it was proved in \cite{torgersen} that
$\Ee\ge_{2,\epsilon} \Fe$ is equivalent with $\Ee\ge_\epsilon \Fe$, so that
for comparison of such experiments it is enough to consider all  testing problems.  
We prove below that this equivalence remains true if
only $\Ee$ is abelian, and that this property characterizes abelian binary
experiments.

We will need the following Lemma.

\begin{lemma}\label{lemma:2points} Let  $s_1,s_2\notin \mathcal T_\Ee$, $0<s_1<s_2$.
Then there is a classical experiment
$\Fe=(X=\{1,2,3\},\{p,q\})$, such that $f_\Ee(t)\ge f_\Fe(t)$ for all $t$ and $f_\Ee(s_i)=f_{\Fe}(s_i)$, $i=1,2$.

\end{lemma}

{\it Proof.} Let us define linear functions $g_i(t):=a_i-tb_i$, $i=0,\dots,3$, where $a_0=b_0=1$, $a_3=b_3=0$  and
$a_i=f(s_i)-s_if'(s_i)$, $b_i=-f'(s_i)$, $i=1,2$, so that
\[
g_i(t)=f_\Ee(s_i)+(t-s_i)f_\Ee'(s_i)
\]
is  tangent to $f_\Ee$ at $s_i$, $i=0,1,2$, where we put $s_0=0$. Since $f_\Ee$ is convex and $f_\Ee(t)\ge \max\{1-t,0\}$,
$g_i(t)\le f(t), $
for all $i$ and $t$. Moreover, since $f_\Ee$ is also  nonincreasing,
we have for any $t<0$, $-1=f_\Ee'(t)\le f_\Ee'(s_1)\le f'_\Ee(s_2)\le 0$ so that
 $b_0\ge b_1\ge b_2\ge b_3$. Convexity and $f_\Ee(0)=1$ also imply that
\begin{eqnarray*}
1-a_1&=&1-f_\Ee(s_1)+s_1f_\Ee'(s_1)\ge 0\\
a_1-a_2&=&f_\Ee(s_1)-f_\Ee(s_2)-f_\Ee'(s_2)(s_1-s_2)+s_1(b_1-b_2)\ge 0\\
a_2&=& f_\Ee(s_2)+s_2b_2\ge 0
\end{eqnarray*}
so that  $a_0\ge a_1\ge a_2\ge a_3$.  Put $p_i:=a_{i-1}-a_i$, $q_i:=b_{i-1}-b_i$,
 $i=1,2,3$, then $p=(p_1,p_2,p_3)$ and $q=(q_1,q_2,q_3)$ are probability measures. Let $\Fe:=(\{1,2,3\},\{p,q\})$, then
\[
 f_\Fe(t)=\sum_{i,p_i-tq_i>0}p_i-tq_i=\sum_{i, g_{i-1}(t)>g_i(t)} g_{i-1}(t)-g_i(t).
\]

Let us now define the points $t'_0,\dots,t'_3$ as follows. Put $t'_0:=0$ and for
$i=1,2,3$, let $t'_i:=t'_{i-1}$ if $g_i=g_{i-1}$, otherwise let
$t_i'$ be such that $g_i(t)<g_{i-1}(t)$
 for $t<t_i'$ and $g_i(t'_i)=g_{i-1}(t'_i)$.  Note that $t'_i\ge 0$, since
 $g_i(0)\le g_{i-1}(0)$. Moreover, since $g_i(s_i)=f_\Ee(s_i)\ge g_{i-1}(s_i)$,
 we have $t'_i\le s_i$ for $i=0, 1,2$. In fact, $t'_i<s_i$ for $i=1,2$, since
 $g_{i-1}(s_i)=g_i(s_i)=f_\Ee(s_i)$ implies $f_\Ee=g_i=g_{i-1}$ in some interval containing
 $s_i$, so that  $t'_i=t'_{i-1}\le s_{i-1}<s_i$. Similarly,
 for $i=2,3$,   $g_i(s_{i-1})\le f_\Ee(s_{i-1})=g_{i-1}(s_{i-1})$, so that we either
 have $t'_i=t'_{i-1}$ or $t'_i>s_{i-1}$. In the case that $g_2(t)>0$ for all $t$, we put $t'_3=\infty$. Putting all together,
 we have $0=t'_0\le t'_1<s_1< t'_2<s_2< t_3'\le\infty$ and
\begin{eqnarray*}
f_\Fe(t)&=&\sum_{j=i}^3g_{j-1}(t)-g_j(t)=g_{i-1}(t),\  t\in \<t'_{i-1},t_i\>, \qquad i=1,2,3\\
f_\Fe(t)&=&0,\ t\in \<t_3,\infty)
\end{eqnarray*}
It follows that $f_{\Fe}(t)\le f_\Ee(t)$ for all $t$ and $f_\Fe(s_i)=f_\Ee(s_i)$, $i=1,2$.

\qed

We will now state the main result of this section.

\begin{theorem}\label{thm:binary_equiv} Let $\Ee=\{\Ae,\{\rho_1,\rho_2\})$ be a binary experiment. Then the following are equivalent.
\begin{enumerate}
\item[(i)] $\Ee\ge_{2,\epsilon}\Fe \iff \Ee\ge_\epsilon \Fe$ for any $\epsilon \ge 0$ and any abelian binary experiment $\Fe$

\item[(ii)] $\Ee\ge_{2,\epsilon}\Fe \iff \Ee\ge_\epsilon \Fe$ for any $\epsilon\ge 0$ and any binary experiment $\Fe$.

\item[(iii)] $\Ee\ge_{2,0}\Fe \iff \Ee\ge_0 \Fe$ for any abelian  binary experiment $\Fe$.
\item[(iv)] $\Ee$ is abelian.

\end{enumerate}

\end{theorem}

{\it Proof.} Suppose (i) and let $\Fe=(\Be,\{\sigma_1,\sigma_2\})$ be any binary
 experiment such that $\Ee\ge_{2,\epsilon} \Fe$. Let $D$ be a finite set and let
$N\in \Me(D,\Fe)$. Put $p_i:=N(\sigma_i)$, $i=1,2$ and let $\Fe_N:=(D,\{p_1,p_2\})$. Then by Theorem \ref{thm:binary}, we have for each $t\ge 0$,
\[
\|\rho_1-t\rho_2\|_1 \ge \|\sigma_1-t\sigma_2\|_1-2(1+t)\epsilon\ge
\|p_1-tp_2\|_1-2(1+t)\epsilon
\]
Hence $\Ee\ge_{2,\epsilon}\Fe_N$ and (i) implies that $\Ee\ge_\epsilon \Fe_N$.
By Corollary \ref{coro:abelian}, there is some $M\in \Me(D,\Ee)$ such that
\[
\|M(\rho_i)-N(\sigma_i)\|_1=\|M(\rho_i)-p_i\|_1\le 2\epsilon,\quad i=1,2
\]
By Theorem \ref{thm:equiv}, $\Ee\ge_\epsilon \Fe$ and this  implies (ii).
(ii) trivially implies (iii).

Suppose (iii). Choose
any points $s_1,s_2\notin \mathcal T_\Ee$,  $0<s_1<s_2$, then by Lemma
\ref{lemma:2points}, there is a classical experiment $\Fe=(\{1,2,3\},
\{p_1,p_2\})$ such that $f_\Ee(t)\ge f_\Fe(t)$ for $t\ge 0$ and $f_\Ee(s_i)=f_\Fe(s_i)$, $i=1,2$. By Theorem \ref{thm:binary}, this implies that $\Ee\ge_{2,0}\Fe$ and
by (iii), $\Ee\ge_0\Fe$. By Corollary \ref{coro:abelian}, there is a POVM
$M: \{1,2,3\}\to \Ae$ such that $p_k=M(\rho_k)$, $k=1,2$. For $i=1,2$, put
$J_i:=\{j\in \{1,2,3\}, p_1(j)-s_ip_2(j)>0\}$, then we have
\begin{eqnarray*}
f_\Ee(s_i)&=&f_\Fe(s_i)=\sum_{j\in J_i}p_1(j)-s_ip_2(j)\\
&=&
 \sum_{j\in J_i}\Tr(\rho_1 M_j)-s_i \Tr( \rho_2 M_j)=
 \Tr (\rho_1-s_i\rho_2) \sum_{j\in J_i}M_j
\end{eqnarray*}
Since $s_i\notin \mathcal T_\Ee$, we have $P_{s_i,0}=0$ and Lemma \ref{lemma:qnp} implies that
$\sum_{j\in J_i}M_j= P_{s_i,+}$. Hence the projection $P_{s_i,+}$ is in the range of $M$. Since for all $j\in \{1,2,3\}$ we either have $M_j\le P_{s_i,+}$ or
$M_j\le I-P_{s_i,+}$, $P_{s_i,+}$ must commute with all $M_j$. In particular,
$P_{s_1,+}$ and $P_{s_2,+}$ commute.

Since this can be done for any such $s_1$, $s_2$, it follows that all
$\{P_{t,+}, t\notin \mathcal T_\Ee \}$ are mutually commuting
projections. Since $t\mapsto P_{t,+}$ is right-continuous,  it follows that
$P_{t_j,+}$ commutes with all $P_{s,+}$ for $s\notin \mathcal T_\Ee$, and by
repeating this argument, $P_{t,+}$ are mutually commuting projections for all
$t\ge 0$.

Let now $\Ae_0$ be the subalgebra generated by $\{P_{t,+}, t\ge 0\}$. Then
$\Ae_0$ is an abelian subalgebra which is  2-sufficient for $\Ee$. Hence  $\Ee$ must be abelian by Theorem \ref{thm:ja_hypo}.

The implication (iv) $\implies$ (i) was proved by Torgersen, \cite{torgersen}.

\qed

\begin{rem} If $\dim(\Ha)=\dim(\Ka)=2$, it was proved in \cite{uhl} that $\Ee\ge_{2,0}\Fe$ if and only if
$\Fe$ is a randomization of $\Ee$. The above proof shows that if $\dim (\Ka)\ge 3$ this is no longer true unless
 $\Ee$ is abelian.

\end{rem}

\section{Statistical morphisms}

Let $S_\Ee:={\rm span}\{\rho_\theta,\theta\in \Theta\}$.
A \emph{$k$-statistical morphism} \cite{buscemi, matsumoto} is  a linear map $L:S_\Ee\to \Be$ such
that
\begin{enumerate}
\item [(i)] $L(\rho_\theta)\in \mathcal S(\Be)$ for all $\theta$
\item[(ii)]  for each POVM $N:D_k\to \Be$ there is some $M\in \Me(D_k,\Ee)$ satisfying
\[
\Tr L(\rho)N_i=\Tr \rho M_i, \qquad i\in D_k,\quad \rho\in S_\Ee.
\]
\end{enumerate}
The map $L$ is a \emph{statistical morphism} if it is a $k$-statistical morphism for any $k$.
It is clear that any positive trace preserving map $L:\Ae\to \Be$  defines a
statistical morphism. The proof of the following proposition appears also in \cite{matsumoto}.

\begin{prop} $\Ee\ge_{k,0}\Fe$ if and only if there is a $k$-statistical morphism $L:S_\Ee\to \Be$ such that $L(\rho_\theta)=\sigma_\theta$.

\end{prop}

{\it Proof.} Suppose that $\Ee\ge_{k,0}\Fe$ for some $k$, then we also have $\Ee\ge_{2,0}\Fe$, and by Theorem \ref{thm:testing},
this implies $\|\sum_\theta A_\theta\rho_\theta\|_1\ge \|\sum_\theta A_\theta\sigma_\theta\|_1$ for any $A_\theta\in \mathbb R$. Put $L:\rho_\theta\mapsto\sigma_\theta$
and extend to $S_\Ee$ by $L(\sum_\theta a_\theta \rho_\theta)=\sum_\theta a_\theta L(\rho_\theta)$,
 then $\|L(x)\|_1\le\|x\|_1 $  for $x\in S_\Ee$, so that $L$  is a well defined   linear map on $S_\Ee$. Theorem \ref{thm:equiv} (iv) now implies
 that $L$ is a $k$-statistical morphism. The converse is obvious.

\qed

In \cite{shmaya} and \cite{buscemi}, a question was raised whether 0-deficiency is equivalent with existence of a 
trace preserving positive map that maps one experiment onto the other. It is clear that this question is equivalent with the
 question if any statistical morphism can be extended to a trace preserving positive map.
     We show below that if $\Ee$ and $\Fe$ are binary experiments, then
any $k$-statistical morphism such that $L(\rho_i)=\sigma_i$, $i=1,2$
can be extended to even a   completely positive map, but Theorem \ref{thm:binary_equiv} implies that such an extension is not trace 
preserving in general. This shows that the condition that the map preserves trace cannot be omitted.

Let $t_1$ be as in Lemma \ref{lemma:fe}. Note that
\begin{equation}\label{eq:t1}
t_1=\max\{t\ge 0, f_\Ee(t)=1-t\}=\max\{t\ge 0, \rho_1-t\rho_2\ge 0\}
\end{equation}
and $t_1=0$ if and only if $\supp \rho_2\not\le\supp \rho_1$.
Let us denote
\begin{equation}\label{eq:tmax}
t_{max}:=\min \{t\ge 0, f_\Ee(t)=0\}=\min\{t\ge 0, \rho_1-t\rho_2\le 0\}.
\end{equation}
Then we have either
$t_{max}=t_l$ or $t_{max}=\infty$, and the latter  happens if and only if $\supp \rho_1\not\le \supp \rho_2$.
We have
\begin{equation}\label{eq:le}
t_1\rho_2\le \rho_1\le t_{max}\rho_2
\end{equation}
and $t_1$, $t_{max}$ are extremal values for which the inequality occurs. Equivalently,
\begin{equation}\label{eq:le2}
t_{max}^{-1}\rho_1\le \rho_2\le t_1^{-1}\rho_1
\end{equation}
with $t_{max}^{-1}$ and $t_1^{-1}$ extremal.
We also remark that $t_1=\sup(\rho_1/\rho_2)$ and $t_{max}=\inf(\rho_1/\rho_2)$ as defined in \cite{wolf}.

\begin{theorem} Let $\Ee=(\Ae,\{\rho_1,\rho_2\})$, $\Fe=(\Be,\{\sigma_1,\sigma_2\})$
be binary experiments. Then if $\Ee\ge_{2,0}\Fe$, then  there is a completely positive map $T:\Ae\to \Be$ such that
$T(\rho_i)=\sigma_i$, $i=1,2$.
\end{theorem}

{\it Proof.} Let  $\Ee\ge_{0,2}\Fe$, then  there is a 2-statistical morphism
$L: S_\Ee\to \Be$, $L(\rho_i)=\sigma_i$, $i=1,2$. Moreover,   $f_\Ee(t)\ge f_\Fe(t)$ for all $t$.
Let $t_1'$ and $t'_{max}$ be as in (\ref{eq:t1}) and (\ref{eq:tmax}) for $\Fe$.
Since $f_\Fe(t)\ge \max\{0,1-t\}$, we must have  $t_1\le t_1'$ and $t_{max}'\le t_{max}$. The rest of the proof is  the same as the
proof of \cite[Theorem 21]{wolf}:

Let $u,v\in S_\Ee$ be positive elements such that ${\rm ker} (u)\not\le{\rm ker}(v)$ and ${\rm \ker}(v)\not\le {\rm ker}(u)$.
Then there are some $\varphi,\psi\in \Ha$
 such that $u\varphi=v\psi=0$, but $u\psi\ne 0$, $v\varphi\ne 0$. Put
 \[
T(a)=\frac{\<\psi,a\psi\>}{\<\psi,u\psi\>}L(u)+\frac{\<\varphi,a\varphi\>}{\<\varphi,v\varphi\>}L(v),\quad a\in \Ae
 \]
then $T$ is a completely positive extension of $L$. We show that such $u$ and $v$ exist.

Suppose $t_{max}<\infty$ so that $\supp \rho_1\le\supp\rho_2$, then $u:=t_{max}\rho_2-\rho_1$, $v:=\rho_1-t_1\rho_2$. Then $u,v\ge 0$ and
the condition on the kernels follows by extremality of $t_1$ and $t_{max}$.
If $t_{max}=\infty$ but $t_1>0$, then we put $u:=t_1^{-1}\rho_1-\rho_2$ and $v:=\rho_2$.
Finally, if $t_{max}=\infty$ and $t_1=0$, then we put $u:=\rho_1$, $v:=\rho_2$.

\qed

\begin{remark} One can see  that the extension
  obtained in the above proof  cannot be trace preserving unless
  $\dim \Ha=2$ and $\Ee$ is abelian.
\end{remark}

\section*{Acknowledgement}

This work was supported by  the grants 
VEGA 2/0032/09 and meta-QUTE ITMS 26240120022.

\section*{Appendix: Proof of Lemma \ref{lemma:fe}.}
The statement (i) follows easily by definition  and (\ref{eq:fe}).

 Let  $\rho(t):=\rho_1-t\rho_2$. It can be shown (\cite[Chap. II]{kato}) that the eigenvalues of 
$\rho(t)$ are analytic functions $t\mapsto \lambda_i(t)$ for all $t\in \mathbb R$. It follows that $\rho(t)$ has a constant number $N$ of distinct eigenvalues
$\lambda_1(t),\dots,\lambda_N(t)$, 
apart from some exceptional points where some of these eigenvalues are equal, 
and there is a finite number of such points in any finite interval. Moreover, let
 $P_i(t)$ be the eigenprojection corresponding to $\lambda_i(t)$ for a 
non-exceptional point $t$, then $t\mapsto P_i(t)$ can be extended to an analytic 
 function for all $t$ such that,  if $s$ is an exceptional point, then the projection corresponding to $\lambda_i(s)$ is given by $\sum_{j,\lambda_j(s)=\lambda_i(s)}P_j(s)$. By continuity,  $\Tr P_i(t)$ is a constant, we denote it by $m_i$. 
If $s$ is not
an exceptional point, $m_i$ is the multiplicity of $\lambda_i(s)$.

By differentiating the equation $\Tr \rho(s)P_i(s)=m_i\lambda_i(s)$ one obtains
\begin{equation}\label{eq:lambda}
\lambda_i'(s)=-\frac 1{m_i}\Tr \rho_2 P_i(s)
\end{equation}
It follows that 
 $\lambda_i(s)$ is nonincreasing  for all $s$,
 moreover, $\lambda_i'(s)=0$ implies that $\rho_2 P_i(s)=0$, so that
 $\rho(t)P_i(s)=\rho(s)P_i(s)=\lambda_i(s)P_i(s)$ for all $t$ and
 $\lambda_i(s)$ is an eigenvalue of $\rho(t)$ for all $t$. Hence
 $\lambda_i$ is either strictly decreasing  or  a  constant, which must be nonzero,
 since we assumed that $\rho_1+\rho_2$ is invertible. It follows that each $\lambda_i$
  hits 0 at most once, so that there is only $l\le N$ points where
  $\lambda_i(t)=0$ for some $i$. Let us denote the points by
  $0\le t_1<\dots<t_l$, it is clear that these are exactly the points where 
$P_{t,0}\ne 0$.
  Let $J_j:=\{i, \lambda_i(t_j)>0\}$, $j=1,\dots,l$. Then  $J_j\subset J_{j-1}$ 
and
\[
f_\Ee(t)=\sum_{i\in J_j}m_i\lambda_i(t),\qquad  t\in \<t_{j-1}, t_j\>,\ 
j=1,\dots,l.
\]
This implies (ii).

\qed

\end{document}